\documentclass[twoside]{amsart}

\usepackage[hmargin={3cm,3cm},vmargin={3cm,3cm},includehead]{geometry}
\usepackage{amsmath,amssymb}
\usepackage{amsfonts}
\usepackage[mathscr]{eucal}
\usepackage{dsfont}

\newcommand{\ds}{\displaystyle}
\def\EXP{\textrm{{\large e}}}

\newcommand{\xop}{{\mathbf{B}}}
\newcommand{\yop}{{\mathbf{B}^\dagger}}

\newcommand{\ii}{\mathsf{i}}

\newcommand{\Nop}{\mathbf{N}}

\newcommand{\density}{\textrm{\LARGE $\varrho$}}
\begin{document}

\vspace{2cm}

\title[]{Evolution operators for quantum chains}%
\author{S. Sergeev}%
\address{Department of Theoretical Physics (Research School of Physical Sciences and Engineering) \& Mathematical Science
Institute, Canberra ACT 0200, Australia}
\email{sergey.sergeev@anu.edu.au}

\thanks{This work was supported by the Australian Research Council}%


\begin{abstract}
Discrete-time evolution operators in integrable quantum lattice
models are sometimes more fundamental objects then Hamiltonians.
In this paper we study an evolution operator for the
one-dimensional integrable $q$-deformed Bose gas with XXZ-type
impurities and find its spectrum. Evolution operators give a new
interpretation of known integrable systems, for instance our
system describes apparently a simplest laser with a clear
resonance peak in the spectrum.
\end{abstract}

\maketitle

Existence of a complete set of commuting operators is the basic
principle of quantum integrability. However, often the commuting
set does not provide a distinguished operator deserving the title
of the Hamiltonian for a physical system. In particular, this is
the common feature of quantum models obtained by quantization of
classical equations of motion in wholly discrete space-time.
Equations of motion define a discrete time translations for
classical variables, a map $A(\tau,\sigma)\to A(\tau+1,\sigma)$.
Corresponding translation for quantum observables is produced by
an evolution operator,
\begin{equation}
\mathbf{A}(\tau+1,\sigma) \;=\; \mathbf{U} \mathbf{A}(\tau,\sigma)
\mathbf{U}^{\dagger}\;, \quad \mathbf{U} \mathbf{U}^\dagger
\;=\;1\;.
\end{equation}
If the time unity interval coincides with the unity spacing, then
there is no a small parameter expansion of $\mathbf{U}$ defining a
\emph{lattice} Hamiltonian. The evolution operator becomes the
main object of the lattice quantum mechanics.

Discrete-time evolution operators for quantum chains were
considered in many papers. For instance, the study of the spectrum
of evolution operator was in the focus of the quantum Liouville
theory \cite{Liouville}. In this paper we discuss another example
of quantum evolution system: the one-dimensional $q$-deformed Bose
gas \cite{Bogolyubov} with $X\!\!X\!\!Z$-type impurities. The
simple test ``what evolution operator is doing in the space of
states'' will give a new view on well known models.

Quantum mechanics begins with the algebra of observables. In our
case the local algebra of observables is the $q$-oscillator
\cite{Kulish} with generators $\xop,\yop,\Nop$:
\begin{equation}\label{q-osc}
\xop\yop=1-q^{2\Nop+2}\;,\quad \yop\xop = 1-q^{2\Nop}\;,\quad 0 <
q < 1\;.
\end{equation}
Below we use the Fock space representation of (\ref{q-osc}), the
Fock vacuum is defined by $\xop|0\rangle = 0$. The $q$-oscillators
are the algebra of observables for $q$-deformed Bose gas. In
addition, it is very convenient to use the $q$-oscillator
representation for $X\!\!X\!\!Z$ model \cite{old,zte}. In this
paper we consider the combined chain of three types of
$q$-oscillators,
\begin{equation}
\{\xop_{j,n},\ \yop_{j,n},\ \Nop_{j,n}\}\;\;:\;\;j=1,2,3\;,\quad
n=0,1,\dots,N-1\;,
\end{equation}
where index ${}_{3,n}$ stands for the $q$-Bose gas, indices
${}_{1,n}$ and ${}_{2,n}$ stand for oscillator representation of
$X\!\!X\!\!Z$ model in $n^{\textrm{th}}$ site of the combined
chain. The quantum Lax operators are
\begin{equation}\label{Lax-boson}
M_n(u)\;=\;\left(\begin{array}{cc} \ds -q^{\Nop_{3,n}}
& \ds u\yop_{3,n}\\
\ds q^{-1}\xop_{3,n} & \ds u q^{\Nop_{3,n}}
\end{array}\right)
\end{equation}
for the $q$-Bose gas (condition $q<1$ is equivalent to the
attractive potential), and
\begin{equation}\label{Lax-spin}
L_n(u)\;=\; \left(\begin{array}{cc} \ds uq^{\Nop_{1,n}} +
\EXP^{\ii\varepsilon} q^{\Nop_{2,n}} & \ds
-u\EXP^{\ii\varepsilon} \xop_{1,n}\yop_{2,n} \\
\ds q^{-1}\xop_{2,n} \yop_{1,n} & \ds u q^{\Nop_{2,n}} +
\EXP^{\ii\varepsilon} q^{\Nop_{1,n}}
\end{array}
\right)
\end{equation}
for the $X\!\!X\!\!Z$ model. $L$-operator (\ref{Lax-spin}) has the
center
\begin{equation}\label{charge}
\mathbf{s}_n\;=\;\frac{1}{2} (\Nop_{1,n} + \Nop_{2,n} )
\end{equation}
with the half-integer spectrum, matrix (\ref{Lax-spin}) on the
subspace $\mathbf{s}_n=s_n$ is identical to the usual spin-$s_n$
$L$-operator for $\mathcal{U}_q(\widehat{sl}_2)$. Note, on the
subspace $\mathbf{s}_n=0$ operator (\ref{Lax-spin}) is just unity
matrix, $\ds L_n(u)|_{\mathbf{s}_n=0}=(u+\EXP^{\ii\varepsilon})\
\mathds{1}$. Extra parameter $\varepsilon$ in (\ref{Lax-spin}) is
equivalent to an exponential shift of the spectral parameter $u$.

The transfer matrix of the model is defined by
\begin{equation}\label{chain}
t(u)\;=\;\textrm{Trace} \biggl( M_0(u) L_0(u) M_1(u) L_1(u) M_2(u)
L_2(u) \cdots M_{N-1}(u) L_{N-1}(u)\biggr)\;.
\end{equation}
Complete set of commutative operators is given by the
decomposition of $t(u)$ and by the set of charges $\mathbf{s}_n$
(\ref{charge}). Subjects of our interest are sub-sectors of the
extended Hilbert space corresponding to specified eigenvalues of
$\mathbf{s}_n$. In particular, if all $\mathbf{s}_n=0$, this is
just the pure one-dimensional $q$-Bose gas with attractive
potential. If $\mathbf{s}_0=1/2$ and all other $\mathbf{s}_n=0$,
this is the Bose gas with the single impurity. If some of
$\mathbf{s}_n=1/2$ and all the other $\mathbf{s}_n=0$ (our final
case), this is the $q$-Bose gas with some density of impurities.

Our main object, the discrete-time evolution operator, is defined
by
\begin{equation}\label{evolution}
\mathbf{U} M_{n}(u) L_n(u) \;=\; L_n(u) M_{n+1}(u) \mathbf{U}\ .
\end{equation}
Evidently, transfer matrix (\ref{chain}) commutes with the
evolution operator and therefore gives the integrals of motion.
One can rewrite (\ref{evolution}) as the map:
\begin{equation}\label{map}
\begin{array}{l}
\ds \mathbf{U}\ q^{\Nop_{2,n}} \yop_{1,n} \mathbf{U}^{-1} =
q^{\Nop_{3,n}}\yop_{1,n} - \EXP^{\ii\varepsilon}
q^{\Nop_{1,n}}\yop_{2,n}\xop_{3,n}\;,\\
\\
\ds \mathbf{U}\ \yop_{2,n} \mathbf{U}^{-1} = \yop_{1,n}\yop_{3,n}
+ q\EXP^{\ii\varepsilon}\
q^{\Nop_{1,n}+\Nop_{3,n}}\yop_{2,n}\;,\\
\\
\ds \mathbf{U}\ q^{\Nop_{2,n+1}} \yop_{3,n} \mathbf{U}^{-1} =
q^{\Nop_{1,n+1}}\yop_{3,n+1} - \EXP^{\ii\varepsilon}
q^{\Nop_{3,n+1}}\xop_{1,n+1}\yop_{2,n+1}\;.
\end{array}
\end{equation}
Corresponding relations for $\mathbf{U} \xop_{j,n}
\mathbf{U}^{-1}$ are the conjugation of (\ref{map}): $\mathbf{U}$
is the unitary operator if $q$ and $\varepsilon$ are real.
Equations (\ref{map}) provide in addition $\mathbf{U} \mathbf{s}_n
= \mathbf{s}_n \mathbf{U}$ and $\mathbf{U} \mathbf{K} = \mathbf{K}
\mathbf{U}$ where
\begin{equation}\label{K}
\mathbf{K} = \sum_{n}\ \left( \Nop_{2,n}+\Nop_{3,n}\right)\;.
\end{equation}
Note, the map (\ref{map}) is exactly the tetrahedral map from
\cite{zte}. All two-dimensional models have hidden
three-dimensional structure.

Take up now the interpretation of the evolution operator. In what
follows, we use
\begin{equation}\label{q-gamma}
q  = \EXP^{-\gamma}\;.
\end{equation}
We start with the lowest occupation numbers over the total Fock
vacuum $|0\rangle$. Let $\mathbf{U}|0\rangle=|0\rangle$, and we
consider next the subspace of the Hilbert space with
$\mathbf{s}_0=\mathbf{K}=1$. Let us see, what the evolution
operator makes with the state
\begin{equation}\label{psi1}
|\psi_1\rangle \;=\; \yop_{2,0}|0\rangle\;.
\end{equation}
Applying (\ref{map}) once, we get
\begin{equation}\label{onestep}
\mathbf{U}\ |\psi_1\rangle \;=\; \EXP^{\ii\varepsilon-\gamma}\
|\psi_1\rangle + |\psi_0\rangle \otimes |b_0\rangle\;,
\end{equation}
where we use the notation
\begin{equation}
|\psi_0\rangle\otimes |b_k\rangle \;=\; \yop_{1,0} \yop_{3,k}
|0\rangle\;.
\end{equation}
What is happening after $\tau$ steps of discrete evolution:
\begin{equation}\label{radiation}
\mathbf{U}^\tau \ |\psi_1\rangle \;=\;
\EXP^{(\ii\varepsilon-\gamma)\tau}\ |\psi_1\rangle \;+\;
\EXP^{-(\ii\varepsilon-\gamma)}\ |\psi_0\rangle\otimes
\left(\sum_{k=0}^{\tau-1} \EXP^{(\ii\varepsilon-\gamma)(\tau-k)}
|b_k\rangle\right)\;.
\end{equation}
Suppose, $\tau$ is big enough, but
\begin{equation}\label{tau-regime}
\gamma\tau \ll 1\;,\quad \varepsilon\tau \simeq \textrm{some
integer multiple of $2\pi$}\;.
\end{equation}
Then equation (\ref{radiation}) describes the decay the
quasi-stable state $|\psi_1\rangle$ with the energy $\varepsilon$
and width $\gamma$. The second term in (\ref{radiation}) is a
state with the bosonic wave function $\Psi(\tau,k)=\EXP^{(\ii
\varepsilon -\gamma)(\tau-k)}$ -- the speed-of-light right-moving
wave radiated by excited state. Due to (\ref{tau-regime}), the
oscillating term in the wave function dominates, and therefore the
wave package in the right hand side of (\ref{radiation}) is the
state with the average energy and momentum $\varepsilon$. Note, we
may talk about the energy of the radiated state only if $\tau$ in
(\ref{radiation}) is big enough, this exactly corresponds to the
Heisenberg uncertainty principle. The one-step distance of the
lattice may be considered as the \emph{ultra-microscopic} scale of
the space-time, while big $\tau$ in (\ref{radiation}) corresponds
to a microscopic scale.

Thus, the discrete-time evolution operator provides an alternative
terminology. We will call site $n$ of the spin chain with
$s_n=1/2$ as the \emph{atom} with two energy states: the ground
state $|\psi_0\rangle$ with zero energy and the excited state
$|\psi_1\rangle$ with the energy $\varepsilon$ and width $\gamma$.
The regime for $\gamma$ and $\varepsilon$ follows from
(\ref{tau-regime}):
\begin{equation}\label{regime}
\gamma \ll \varepsilon \ll 2\pi\;.
\end{equation}
The spin chain sites with $s_n=0$ do not provide any effect, these
sites are just empty. The Bose field $\yop_{3,n}$ is identified
with the electro-magnetic field, the radiated state in
(\ref{radiation}) is the \emph{photon}. The whole system is
evidently a kind of chiral laser (photons are radiated to the
right only). Note, the impurities destroy the translation
invariance of the system. Scattered and radiated photons may have
momenta, but the atoms are rigidly fixed and the solid media takes
up the recoil momenta of photons.

Consider next eigenstates and eigenvalues of the evolution
operator. We consider the periodical boundary conditions -- a
``toroidal chiral laser''. The eigenstates have the structure of
the superposition of stationary bosonic waves scattering on the
atoms (the coordinate Bethe Ansatz). For instance, the one-boson
and one-atom eigenstate $|\Psi_p\rangle$, such that $\mathbf{U}
|\Psi_p\rangle = \EXP^{\ii p} |\Psi_p\rangle$, is
\begin{equation}\label{one-state}
|\Psi_p\rangle \;=\;
-\frac{\EXP^{\ii(p+\varepsilon)}(1-q^2)}{q\EXP^{\ii
p}-\EXP^{\ii\varepsilon}} \yop_{2,0}|0\rangle \;+\;\yop_{1,0}
\yop_p |0\rangle\;,
\end{equation}
where
\begin{equation}\label{photon}
\yop_p=\sum \EXP^{-\ii p \, k} \yop_{3,k}
\end{equation}
is the creation operator of the photon. The quantization equation
for the momenta of one-photon states follows from the periodical
boundary conditions,
\begin{equation}\label{bae-11}
\EXP^{\ii N p }\ \frac{\EXP^{\ii p} - q\EXP^{\ii
\varepsilon}}{q\EXP^{\ii p} - \EXP^{\ii \varepsilon}}\;=\; 1\;.
\end{equation}
Since the evolution operator describes only the right-moving
photons with dispersion relation $E=p$, the momentum $p$ in
(\ref{photon},\ref{bae-11}) must be positive.

In general, the spectrum of $\mathbf{U}$ is given by
\begin{equation}\label{U-eigen}
\mathbf{U}\;=\;\prod_{i=1}^K u_i\;=\;\exp\left\{\ii\sum_{i=1}^K
p_{\,i}\right\}\;,
\end{equation}
where $\ds u_i=\EXP^{\ii p_{\,i}}$ are the roots of the Bethe
Ansatz equations
\begin{equation}\label{BAE}
u_i^N\ \prod_{n} \frac{u_i -
q^{2s_n}\EXP^{\ii\varepsilon}}{q^{2s_n}u_i-\EXP^{\ii\varepsilon}}
\;=\; \prod_{j\neq i}
\frac{q^{-1}u_i-qu_j}{qu_i-q^{-1}u_j}\;,\quad i,j=1,...,K\;.
\end{equation}
Here $N$ is the length of the chain, periodical boundary
conditions are taken into account, and $K$ is the eigenvalue of
(\ref{K}) (the number of photons in the laser). In general, $s_n$
are arbitrary eigenvalues of $\mathbf{s}_n$, but here we consider
$s_n=0,1/2$ only. Equations (\ref{BAE}) are literally the Bethe
Ansatz equations for transfer matrix (\ref{chain}); relation
(\ref{U-eigen}) is the main analytical result of this paper.

The eigenstates corresponding to (\ref{BAE}) may be interpreted in
terms of the photon creation operators in momentum space
(\ref{photon}). If all $p_i$ are real and different, the photonic
counterpart of the eigenstate is a slightly modified state
$\yop_{p_1}\yop_{p_2}...\yop_{p_K}|0\rangle$. In addition to real
$p_i$, the Bethe Ansatz equations (\ref{BAE}) have solutions with
complex $p_i$ (strings of the Bethe Ansatz equations), for
instance $p_1\simeq p+\ii\gamma$ and $p_2 \simeq p-\ii\gamma$. The
bosonic counterpart of such eigenstate is $\ds\sum_{k,k'}
\EXP^{-\ii p (k+k') - \gamma |k-k'|}
\yop_{3,k}\yop_{3,k'}|0\rangle$. This is the bound state for big
$\gamma$, but in our regime (\ref{regime}) it may be understood as
properly $\gamma$-regularized  state $\yop_{p}^2|0\rangle$.
Analogously, the highest bound states correspond to the highest
occupation numbers $\yop_{p}^n|0\rangle$. The important point is
that we have to solve the Bethe Ansatz equations implying the
right-moving condition $\textrm{Re}(p_i)>0$.

Turn finally to the thermodynamical limit $N,K\to \infty$. Let the
number of two-states atoms be $N'<N$, their density $n=N'/N$ is
not zero at the limit. Structure of the infrared part of spectrum
is the subject of separate investigation. The lowest energy levels
correspond to Bethe Ansatz strings, the $\gamma$-regularization
solves apparently the infrared catastrophe problem.

In this paper we consider the ``optical wave band'' $p\sim
\varepsilon$. Assume that the Bethe roots $u_i=\EXP^{\ii p_i}$,
$\textrm{Im}(p_i)=0$, form a dense distribution
$\density(p_i)^{-1}=N(p_{i+1}-p_i)$ near $p_i\sim \varepsilon$.
Assume in addition that we can neglect in this region densities of
photons with higher occupation numbers (complex $p_i$). The
standard Bethe Ansatz technique provides the Hulthen integral
equation for the density $\density(p)$:
\begin{equation}\label{tba}
1 - 2\pi \density(p) \;=\; \int dp' \density(p') \frac{d}{dp}
S(p,p') + n \frac{d}{dp} F(p)\;,
\end{equation}
where
\begin{equation}
\EXP^{\ii S(p,p')}\;=\;\frac{\EXP^{\ii p+\gamma}-\EXP^{\ii
p'-\gamma}}{\EXP^{\ii p-\gamma}-\EXP^{\ii p'+\gamma}}\;,\quad
\EXP^{\ii F(p)}\;=\;\frac{\EXP^{\ii
p-\gamma}-\EXP^{\ii\varepsilon}}{\EXP^{\ii p } -
\EXP^{\ii\varepsilon - \gamma}}\;.
\end{equation}
Solving (\ref{tba}) in the optical wave band, we neglect the
contribution of the radio band since $\ds\frac{d}{dp}S(p,p')\to 0$
when $|p-p'| \gg \gamma$. The solution is
\begin{equation}\label{answer}
\density(p)\;\simeq\; (4\pi)^{-1} \;+\;\frac{n}{4\gamma}\
\frac{1}{\cosh \frac{\pi (p-\varepsilon)}{2\gamma}}\;,
\end{equation}
what is evidently the laser resonance peak over the white noise.
Accounting of radio band and complex $p_i$ may modify the white
noise term. The resonance term has exactly the $X\!\!X\!\!X$-spin
chain structure, so that the number of resonant photons in the
stationary state is half of the number of atoms. Additionally
pumped energy does not increase the density of resonant photons,
it dissipates into the white noise.

\noindent{\textbf{Acknowledgements.}} I would like to thank M.
Bortz, J. de Gier  and V. Mangazeev for fruitful discussions.

\end{document}